\documentstyle[12pt,version2,aps]{revtex}
\begin{document}
\setlength{\baselineskip}{3ex}

\renewcommand{\theequation}{\arabic{section}.\arabic{equation}}
\newcommand{\eqreset}{\setcounter{equation}{0}}
\begin{center}
{\large\bf Universal Finite-Size Scaling Function \vspace{0mm}\\
of the Ferromagnetic Heisenberg Chain  \vspace{4mm}\\
in a Magnetic Field\footnote{
{\normalsize Submitted to J.Phys.Soc.Jpn}
}
} \vspace{5mm}\\
 Hiroaki  {\sc Nakamura}\footnote{e-mail "hiroaki@eagle.issp.u-tokyo.ac.jp"} ,
 Naomichi $ {\mbox {\sc Hatano}}^1  $
         and
    Minoru {\sc Takahashi}
 \vspace{5 mm}\\
        {\it Institute for Solid State Physics, University of Tokyo}\\
        {\it Roppongi, Minato-ku, Tokyo 106} \\
        {\it ${}^{ 1}$Department of Physics, University of Tokyo}\\
        {\it Hongo, Bunkyo-ku, Tokyo 113  } \\
\end{center}
\vspace{5 mm}
\begin{abstract}
%
The finite-size scaling function of the  magnetization  of the ferromagnetic
Heisenberg chain  is argued to be universal with respect to the magnitude of
the spin.
The finite-size scaling function is given explicitly
by an analytical calculation
in the classical limit $S=\infty.$
The universality is checked for $S=1/2$ and $1$
by means of numerical calculations.
Critical exponents are obtained as well.
It is concluded
that this universal scaling function originates
in the universal behavior of the correlation function.

\vspace{10mm}
\end{abstract}
\pagebreak
%
\section{Introduction}\label{intro}
\setcounter{equation}{0}
\newcommand{\qbSpin}{\mbox{\boldmath $S$}}
It is a fairly long time since the ground-state criticality of the
ferromagnetic Heisenberg chain was studied first for the classical spin.
To extend the study to the quantum spin system required considerable effort,
though it might look rather easy.
The interest was renewed quite recently when the universality of the system was
explored for any values of $S$.

In the nineteen-sixties, Fisher explicitly calculated\cite{Fisher} the free
energy, the susceptibility and the correlation function  of the classical
Heisenberg chain of length $L$  under the open boundary condition.
According to his results, the susceptibility diverges as $1/T^2$ at $T=0$ in
the ferromagnetic case.
The correlation function decays strictly exponentially,
and the correlation length diverges as $1/T$ at $T=0$.

As for the periodic chain,
Joyce obtained\cite{Joyce} exact expressions for
the partition function,
the correlation function
and the susceptibility  of
the classical Heisenberg chain.
His results agree with those of the open chain in the thermodynamic limit $L
\rightarrow 0,$ as would be expected.

For the quantum system, on the other hand,
Yamada and Takahashi\cite{TY1,TY2,Yamada} employed the Bethe-ansatz method,
and
numerically obtained the free energy and  the zero-field susceptibility of the
$S=1/2$ ferromagnetic Heisenberg chain  in the thermodynamic limit.
Their calculations gave critical exponents as $\alpha = -1/2 ,$
$\nu =1 $ and $\gamma =2 .$
Moreover, they found that the free energy and the susceptibility  can be
expanded in $ \sqrt{T} $ at low temperatures.

Recently, two of the present authors numerically calculated\cite{NT} the free
energy of the $S=1/2$ Heisenberg chain with a magnetic field in the
thermodynamic limit.
We concluded in that work that  the $S=1/2$ ferromagnetic Heisenberg chain  in
the thermodynamic limit has the same scaling function of the magnetization as
that of the classical ferromagnetic Heisenberg chain.
Moreover, we conjectured that this scaling function
should be universal for any values of $S.$

In the present paper,
we find the {\em universal} finite-size scaling function
for the ferromagnetic Heisenberg chains.
Thus we conclude that the ferromagnetic Heisenberg chains with arbitrary $S$
belong to the same universality class.

In \S2,
we review the scaling function and the critical exponents of
the arbitrary-$S$ Heisenberg chain in the thermodynamic limit,
in order to obtain the scaling variable.
We present the finite-size scaling function explicitly in \S3.
We also show numerical confirmation of the universality for $S=1/2$ and $1$.
In \S4,
we argue the origin of the universal scaling function of the susceptibility.

\eqreset
%
\section{Scaling Function in the Thermodynamic Limit}\label{sec.2}
\setcounter{equation}{0}
In this section we discuss critical exponents
and the scaling law of the magnetization
of the arbitrary-$S$
ferromagnetic Heisenberg chain
in the thermodynamic limit.
Thus we know what is the scaling variable for the finite-size scaling function
given in the next section.

The spin-$S$ ferromagnetic Heisenberg chain
is defined as follows:
\begin{equation}
{\cal H} = - \frac{J}{S^2} \sum_{i=1}^{L}
            {\qbSpin}_{i}\cdot {\qbSpin}_{i+1}
           - \frac{h}{S} \sum_{i=1}^{L}
              S_{i}^{z} ,  \label{heisen}
\end{equation}
where
the coupling $J$ is positive,
and $h$ denotes a magnetic field.
We consider both periodic and open boundary conditions.

The magnetization of the
ferromagnetic Heisenberg chain
has the singularity
at $T=h=0,$
where the susceptibility and the correlation length diverge.
At low temperatures,
some physical quantities generally show the following scaling behaviors:
\begin{eqnarray}
m &\sim& T^{\beta} {\tilde m} \left(  \frac{h}{T^\Delta} \right)
, \label{beta} \\
m &\sim& h^{ 1/ \delta } \quad \mbox{at}\quad T=0,\label{delta}\\
\chi &\sim& T^{-\gamma} ,\label{gamm} \\
\xi &\sim& T^{-\nu} , \label{nu}
\end{eqnarray}
where
$ \xi $ is the correlation length
and $ \chi $ is the linear susceptibility.

In the following, we review the scaling relations among the above critical
exponents.
The correlation function $f(r)$ between two spins with a length $r>0$
shows the scaling behavior as\cite{TK1,STK,TK2}
\begin{equation}
f(r) \equiv \left\langle \left( \frac{S_1^z}{S} \right)
 \left( \frac{ S_{1+r}^z }{S} \right)
  \right\rangle \sim
 { \left( \frac{T}{T_0} \right) }^{ 2 \beta} \phi \left( r/\xi \right),
\label{eta}
\end{equation}
where
the angular brackets $ \langle \cdots \rangle $
denote the thermal average, and $T_0 $ is a characteristic temperature.
The scaling function $\phi(x)$ has the following form:\cite{TK1,STK,TK2}
\begin{eqnarray}
\phi(x) &=& \frac{ {\rm e}^{-x} }{ x^{\eta} } \nonumber \\
        &\sim&
          \left\{
        \begin{array}{ll}
         {\rm e}^{ - x}       &  \mbox{ for } x \gg 1  , \\
         \frac{1}{ x^{\eta} } &  \mbox{ for } x \ll 1  ,
        \end{array}
  \right.   \label{as.corre}
\end{eqnarray}
with the critical exponent $\eta.$
As $T$ decreases to zero,
for $r \ll \xi,$
the correction function $f(r)$ behaves as
\begin{eqnarray}
f(r) &\sim& { \left( \frac{T}{T_0} \right) }^{ 2\beta}
            {\left( \frac{r}{\xi} \right)}^{-\eta}  \label{sc00} \\
     &\sim& T^{ 2\beta-\nu \eta} r^{-\eta}       ,   \label{sc1}
\end{eqnarray}
where we use the scaling behavior of $\xi$, (\ref{nu}).
Since the correlation function $f(r)$ remains finite at $T=0,$
we have the following relation\cite{TK1,STK,TK2} between the critical
exponents:
\begin{equation}
2\beta = \nu \eta . \label{sl1}
\end{equation}
The correlation thereby reduces to
\begin{equation}\label{corrT0}
f(r)\sim r^{-\eta}\quad\mbox{at}\quad T=0.
\end{equation}
Furthermore, we obtain the susceptibility from (\ref{eta}) as\cite{TK1,STK,TK2}
\begin{eqnarray}
\chi &=& \frac{1}{T} \int f(r) {\rm d} r \nonumber \\
     &\sim& \frac{ \xi T^{ 2 \beta} }{T}
           \int \phi (r/\xi) {\rm d} ( r/\xi)  \nonumber \\
     &\sim& T^{2\beta - \nu -1 } 	.	\label{sc2}
\end{eqnarray}
Comparing eqs.(\ref{gamm}) and (\ref{sl1}) with eq.(\ref{sc2}),
we have the following scaling relations:
\begin{equation}
\gamma = -2\beta+\nu +1 = \nu(1-\eta)+1 .\label{sl2}
\end{equation}
Finally, the standard scaling argument is followed by the relations
\begin{eqnarray}
\Delta &=& \beta + \gamma , \label{sl3} \\
\frac{\beta}{\Delta} &=& \frac{1}{\delta} . \label{sl4}
\end{eqnarray}

Now we show the exponent relations for ferromagnetic Heisenberg models with
the $T=0$ singularity.
%
%
First, at $T=0,$
the argument in Appendix yields
\begin{equation}
3\left\langle \left( \frac{S_1^z}{S}  \right)
\left( \frac{ S_{1+r}^z }{S} \right)  \right\rangle
  = \left\langle
        \frac{ {\qbSpin}_{1}\cdot {\qbSpin}_{1+r} }{S^2 }
       \right\rangle
= 1 , \label{2.3}
\end{equation}
where these angular brackets $\langle\cdots\rangle$ denote the expectation
value at the ground state instead of the thermal average.
Comparing eq.(\ref{corrT0}) with eq.(\ref{2.3}), we obtain
\begin{equation}
\eta =0 . \label{2.5}
\end{equation}
This result (\ref{2.5}) is consistent with
Fisher's result\cite{Fisher} for $S=\infty.$
Then the scaling relation (\ref{sl1}), $2\beta=\nu\eta$, gives
\begin{equation}
\beta = 0.\label{2.6}
\end{equation}
Next, at $T=0$ and with a magnetic field,
the system should be fully magnetized; hence
eq.(\ref{delta}) should reduce to
\begin{equation}
m \equiv 1 .\label{2.7}
\end{equation}
Thus we have
\begin{equation}
\frac{1}{\delta} =0, \label{2.8}
\end{equation}
which is consistent with $\beta=0$ through the scaling relation (\ref{sl4}).
Now the scaling relations (\ref{sl2}) and (\ref{sl3})
give
\begin{equation}
\Delta = \gamma = \nu +1 . \label{2.13}
\end{equation}
The above indicates that the $T=0$ singularity of ferromagnetic Heisenberg
models is described by the one-parameter scaling.

We are in a position to obtain all the exponents for the one-dimensional
model.
We proposed\cite{NT} the thermodynamic-limit scaling function for arbitrary
magnitude of the spin $S$ as follows:
\begin{equation}
m(T,h) = \tilde m_0 \left( \frac{ Jh}{T^2} \right) ,  \label{2.1}
\end{equation}
where
\begin{equation}
\tilde m_{0}  (x) = \frac{2}{3} x - \frac{44}{135} x^3
+O \left( x^5 \right) . \label{2.2}
\end{equation}
The scaling function of the susceptibility is hence given by
\begin{equation}
 \chi(T,h) = \frac{J}{T^2} {\tilde \chi_0}
        \left( \frac{Jh}{T^2} \right) \label{sh.chi}
\end{equation}
with
\begin{equation}
 {\tilde \chi_0} (x) = \frac{2}{3} -
                     \frac{44}{45} x^2
                     + O \left(  x^4 \right)  .  \label{sh.cale.chi}
\end{equation}
Comparing (\ref{2.1}) with (\ref{beta}),
we can conclude
\begin{equation}
\Delta = 2
\quad\mbox{and}\quad
\beta = 0 \quad \mbox{for } {}^{ \forall } S . \label{2.16}
\end{equation}
Equations (\ref{2.13}) and (\ref{2.16}) give other critical exponents
for the arbitrary spin
as follows:
\begin{equation}
\nu = 1
\quad\mbox{and}\quad
\gamma =2 \quad \mbox{ for } {}^{ \forall }  S .\label{2.17}
\end{equation}
As for the classical limit $S=\infty,$
Fisher\cite{Fisher}
had obtained the values (\ref{2.17})
by solving the model analytically.
For $S=1/2,$
on the other hand,
Yamada and Takahashi\cite{TY1,TY2,Yamada}
had obtained the values (\ref{2.17}) numerically.

We comment on the critical exponent $ \alpha,$
which denotes the singularity of the specific heat $C$
in the form
$ C \sim T^{ \alpha }.$
In an ordinary case,
we get the following scaling law:
\begin{equation}
 \alpha + 2\beta + \gamma =1 . \label{ascale}
\end{equation}
In the case of $S=1/2$ of our system (\ref{heisen}),
numerical calculations\cite{TY1,TY2} gave $\alpha$ as
\begin{equation}
\alpha = -1/2 . \label{alpha}
\end{equation}
Then the values $\beta =0$
and $\gamma = 2$ in (\ref{2.16}) and (\ref{2.17})
appear to be inconsistent with (\ref{ascale}) and (\ref{alpha}).
This inconsistency
comes from a particular term of the free energy.
According to our numerical calculations,\cite{NT}
the leading term of the free energy,
$T^{3/2},$
does not depend on the magnetic field.
The magnetic properties are governed
by the next term of the free energy,
$T^2.$

%
%
%
%
\section{Finite-Size Scaling Function}\label{sec.3}
\setcounter{equation}{0}
In this section,
we extend the scaling forms in the previous section
to the finite-size system.
We use the dimensionless scaling parameter
$ L/\xi ( = T^{\nu} L )$ as the scaled length.

\subsection{General Consideration}\label{sec.3.1}
The finite-size scaling form near the singularity
$T=h=0$ is generally given\cite{privman} in the form
\begin{equation}
 m(T, h, L) = T^{ \beta }   \tilde m
    \left( \frac{ (h/J) }{ {(T/J)}^{ \Delta} } ,
    \frac{1}{ { (T/J ) }^{\nu} L}
    \right) .     \label{3.1}
\end{equation}
We substitute the critical exponents in eq.(\ref{3.1})
with eqs.(\ref{2.16}) and (\ref{2.17}),
and we have
\begin{equation}
 m(T, h, L) =   \tilde  m
    \left( \frac{Jh}{T^{2} } , \frac{J}{TL}
    \right)   .   \label{3.2}
\end{equation}
In the thermodynamic limit $ L \rightarrow \infty ,$
eq.(\ref{3.2}) becomes $\tilde m_{0} $
given by eq.(\ref{2.2}):
\begin{equation}
m(T,h,L\rightarrow \infty ) \rightarrow \tilde m  \left( \frac{Jh}{T^2 } , 0
\right)
       = \tilde m_0 \left( \frac{ Jh}{T^2} \right) . \label{3.3}
\end{equation}
Since ${\tilde m_0}$ is conjectured
to be universal,\cite{NT}
we expect
that the finite-size scaling function $\tilde  m$
in (\ref{3.1})
is also universal,
or
common to all magnitudes of $S.$

The function (\ref{3.2}) is immediately followed by the scaling function
of the zero-field susceptibility of the form
\begin{equation}
\chi (T,L) \equiv  \left. \frac{ \partial m }{ \partial h }
\right|_{h=0} = \frac{J}{T^2} \tilde \chi \left( \frac{J}{TL} \right) ,
\label{3.2.1}
\end{equation}
where
\begin{equation}
 \tilde \chi\left( y \right) \equiv
          \left. \frac{ \partial {\tilde m (x,y) } }{ \partial x }
          \right|_{x=0} . \label{3.2.2}
\end{equation}
The finite-size scaling function $\tilde \chi$ does not
depend on the magnitude of the spin $S,$
because $ \tilde m$ should be independent of $S.$

\subsection{Analytical calculation in the classical case (S=
$\infty$)}\label{sec.3.2.1}
\newcommand{\cbSpin}{\mbox{\boldmath $n$}}
Here
we consider the classical Heisenberg chain
in order to obtain the finite-size scaling function
$\tilde \chi$ exactly.
The classical Heisenberg model is given as follows:
\begin{eqnarray}
 {\cal H} &=& - \frac{J}{S^2}  \sum_{i=1}^{L} { \qbSpin }_{i}
\cdot { \qbSpin }_{i+1} -   \frac{h}{ S} \sum_{i=1}^{L} { S}_{i}^{z}
\nonumber \\
        &\rightarrow&  - J \sum_{i=1}^{L}
        \cbSpin_{i} \cdot \cbSpin_{i+1}
        - h \sum_{i=1}^{L} n_{i}^{z}  \quad \mbox{as} \quad
S \rightarrow \infty, \label{class}
\end{eqnarray}
where \{ $  {\cbSpin}_{i} $ \} are vectors of length unity.

%
%
First,
we consider the periodic boundary condition.
Joyce\cite{Joyce} obtained
the partition function of this model, ${\cal Z}_L,$
as follows:
\begin{equation}
{\cal Z}_L (K) = \sum_{l=0}^{\infty} (2l+1) \lambda_{l}^{L} (K)
,\label{class.prt}
\end{equation}
where $K \equiv J/T$ and
\begin{equation}
\lambda_{l} (K) \equiv  \sqrt{ \frac{ \pi }{2K} }
\ \ {I}_{l+ \frac{1}{2} } (K) \label{lambda}
\end{equation}
with $ {I}_{l+ \frac{1}{2} } (K) $
being the modified Bessel functions of the first kind.
The correlation function is given by\cite{Joyce}
\begin{equation}
\langle  n_{1}^{z} n_{ 1+M }^{z} \rangle_L^{\rm peri}
       = \frac{1}{ {\cal Z}_L }
         \sum_{l=0}^{\infty} (l+1) \lambda_{l}^{L}
         \left[
            \left( \frac{ \lambda_{l+1} }{ \lambda_{l} } \right)^{M}
            +
            \left( \frac{ \lambda_{l+1} }{ \lambda_{l} } \right)^{L-M}
         \right] , \label{pericorr}
\end{equation}
where the  subscript $L$ denotes the total number of spins.
The summation of the correlation functions (\ref{pericorr})
gives the zero-field linear susceptibility per site,
$\chi,$ in the form
\begin{equation}
3T \chi^{\rm peri} (K,L)
= 1 + \frac{2}{{\cal Z}_L (K) }  \sum_{l=0}^{\infty}  (l+1) \left[
\frac{ \lambda_{l} (K)  \lambda_{l+1}^{L} (K)
- \lambda_{l+1} (K) \lambda_{l}^{L} (K) }
{ \lambda_{l+1} (K) - \lambda_{l} (K) } \right]  . \label{class.chi}
\end{equation}

We are interested in the scaling limit of eq.(\ref{class.chi}).
Noting the scaling form (\ref{3.2.1}), we leave only the contributions of the
scaling variable $TL$ in the limit $T\to0$ and $L\to\infty$.
At low temperatures,
$ \lambda_{l}$ behaves as follows:
\begin{eqnarray}
\lambda_0 (K) &\rightarrow& \frac{ \exp{K} }{2K}
             \quad \mbox{as}\quad K \rightarrow \infty, \label{aslambda0} \\
\left[ \frac{ \lambda_{l+1} (K)}{ \lambda_{l} (K)  } \right]^{N}
         &\rightarrow& \exp{ \left[  - ( l+1 ) \frac{ N}{K} \right] }
                \quad    \mbox{as} \quad K \rightarrow \infty,
\label{aslambda1}
\end{eqnarray}
where $N$ is an arbitrary positive integer.
Hence we have
the susceptibility at low temperatures in the following scaling form:
\begin{equation}
 \chi^{\rm peri}(T,L)  \sim \frac{2J}{3T^2}\left\{ \sum_{l=0}^{\infty} (2l+1)
 \exp \left[ - \frac{ l(l+1) }{2}   \frac{TL}{J} \right] \right\} ^{-1}
 + O\left( \frac{1}{T} \right) . \label{finite_scale1}
\end{equation}
Comparing eq.(\ref{finite_scale1}) with the scaling form eq.(\ref{3.2.1}),
we obtain the finite-size scaling function of the susceptibility
as follows:
\begin{equation}
{\tilde \chi}^{\rm peri} (y) = \frac{2}{3}
\left\{ \sum_{l=0}^{\infty} (2l+1)
 \exp \left[ - \frac{ l(l+1) }{2y}
 \right] \right\}^{-1} .
\label{finite_scale}
\end{equation}
The asymptotic behavior of ${\tilde \chi}^{\rm peri}$ is derived as
\begin{equation}
{\tilde \chi}^{\rm peri}(y) \sim
  \left\{
        \begin{array}{ll}
         \frac{2}{3} \left[ 1-3 \exp{ \left( - \frac{1}{y} \right) } \right]
\rightarrow \frac{2}{3} & \mbox{ as } y \rightarrow 0 , \\
         \frac{1}{3y} \left( 1 - \frac{1}{6y} \right) \rightarrow 0
&  \mbox{ as } y \rightarrow \infty  ,
       \end{array}
  \right.
  \label{asperi}
\end{equation}
The liming behavior in $y \rightarrow 0 ,$
or $L \rightarrow \infty ,$
corresponds to the first term of ${\tilde \chi_0}$
in eq.(\ref{sh.cale.chi}).

We show in Fig.\ref{f1} the temperature dependence
of the linear susceptibility $\chi^{\rm peri}$ with $J/(TL)$ fixed.
In the numerical calculation of eq.(\ref{class.chi}),
we sum up the terms up to $l=100,$
which is large enough  for this summation to converge.
We see the clear convergence to the scaling limit (crosses),
which are given by eq.(\ref{finite_scale}).

%
%
%
Next, we study the open system.
In this case,
Fisher\cite{Fisher} gave the correlation function
analytically as follows:
\begin{equation}
\langle  n_{1}^{z} n_{ 1+M }^{z} \rangle^{\rm open}
 = \frac{1}{3} {u(K)}^M,
             \label{opencorr}
\end{equation}
where
\begin{equation}
u(K)\equiv\coth K-\frac{1}{K}.
\end{equation}
Summing up the correlations, we obtain the linear susceptibility  of the open
system as
\begin{equation}
3T\chi^{\rm open}(K,L)
=\frac{1+u}{1-u}-\frac{2u(1-u^L)}{L(1-u)^2}.
\end{equation}
Taking the scaling limit $T\to0$ and $L\to\infty$ with $TL$ fixed,
we have the scaling form of
the susceptibility in (\ref{3.2.1}) in the form
\begin{equation}
{\tilde \chi}^{\rm open} (y) =
           \frac{2}{3}
           \left\{
             1 - y \left[ 1- \exp{ \left( -\frac{1}{y} \right)  } \right]
           \right\} . \label{open}
\end{equation}
The asymptotic behavior of this scaling function is of the forms
\begin{equation}
{\tilde \chi}^{ \rm {open} } (y) \sim
  \left\{
        \begin{array}{ll}
         \frac{2}{3} (1-y)  \rightarrow \frac{2}{3}
& \mbox{ as } y \rightarrow 0 , \\
         \frac{1}{3y} \rightarrow 0  &  \mbox{ as } y \rightarrow \infty  .
       \end{array}
  \right.
  \label{asopen}
\end{equation}
The value in $y \rightarrow 0,$
or $L\rightarrow \infty ,$
is the same as in (\ref{asperi}).
We show the temperature dependence
of the linear susceptibility (\ref{open})
in Fig.\ref{f5}.
The convergence to the scaling limit is clear in the open system as well.

\subsection{Numerical calculations in the quantum case (S= $1/2$ and
$1$)}\label{sec.3.2.2}
Now that we analytically obtained the finite-size scaling functions
(\ref{finite_scale}) and (\ref{open}),
we confirm them for $S=1/2$ and $1$ by means of numerical calculations.
In order to calculate the susceptibility numerically,
we diagonalized the Hamiltonian exactly by the Householder method.
We treated the systems of size up to $L=14$
for $S=1/2,$
and up to $L=8$ for $S=1$ with both periodic and open boundary conditions.
We calculated the susceptibility as
\begin{equation}
\chi(T,L) = \frac{1}{TL}
\sum_{i=1}^{L} \sum_{j=1}^{L}
\left\langle
\left( \frac{ S_{i}^{z}}{S} \right)
\left( \frac{ S_{j}^{z}}{S} \right)
\right\rangle . \label{num.1}
\end{equation}

First,
the numerical results for the periodic boundary condition
are plotted in Figs.\ref{f2} and \ref{f3}.
In these figures,
we assumed that the leading correction
to the finite-size scaling form is of the form $ \sqrt{T/J} ,$
because the susceptibility for $S=1/2$
can be expanded with respect to $ \sqrt{T/J} $
in the thermodynamic limit.\cite{TY1,TY2,NT}
(This leading correction may vanish in the classical limit
as we observe in Fig.\ref{f1}.)
We extrapolated the scaling limit quadratically,
using the three points nearest to the ordinate.

In Fig.\ref{f4},
we summarize the extrapolated date of Figs.\ref{f2} and \ref{f3}
with the scaling function (\ref{finite_scale}).
The data for $S=1/2$ and $S=1$ are quite consistent with
the finite-size scaling function for $S= \infty.$
This fact strongly suggests that
the finite-size scaling function $ {\tilde \chi} $ does not depend on
the magnitude of the spin.

Next,
for the open boundary condition,
we carried out in Figs.\ref{f6} and \ref{f7} the same analysis
as in Figs.\ref{f2} and \ref{f3}.
The finite-size scaling function (\ref{open})
and the scaling-limit data from Figs.\ref{f6} and \ref{f7}
are shown in Fig.\ref{f8}.
The agreement is remarkable.
We conclude that  the finite-size scaling function (\ref{open}) is common to
all magnitudes of the spin  for the open system as well.

\eqreset
\section{Discussions}\label{sum}
\setcounter{equation}{0}
We showed above that the finite-size scaling function $ \tilde{ \chi} $ is
universal with respect to the magnitude of the
spin both for the periodic and open boundary conditions.
The scaling functions are given by eqs.(\ref{finite_scale}) and (\ref{open}),
respectively.
We checked the universality
for $S=1/2$ and $1$ numerically.

We now naturally expect that this universality of the susceptibility originates
in the  universality of the correlation function.
Then we may obtain the scaling function
of the correlation function for arbitrary $S$
on the basis of the classical limit.
For the periodic boundary condition,
the correlation function behaves in the scaling limit
as follows:
\begin{equation}
 \left\langle
\left( \frac{ S_{1}^{z}}{S} \right)
\left( \frac{ S_{1+M}^{z}}{S} \right)
\right\rangle_L^{\rm peri}
 \sim
    f^{\rm peri} \left(M,\frac{J}{TL} \right) \quad \mbox{ for} {}^{ \forall }
S .\label{pcor}
\end{equation}
By using the classical correlation function (\ref{pericorr}),
we obtain the scaling function
$f^{\rm peri}(M,J/TL)$ for the periodic system
as follows:
\begin{equation}
f^{\rm peri}(M,y) \equiv
          \frac{
          \sum_{l=0}^{\infty}
            \left\{ (l+1) \exp{ \left[ - \frac{l(l+1)}{2y}
  \right] } \right\}
                        \left[
                          \exp{ ( -\frac{M}{ \xi_l} ) }
+  \exp{ ( -\frac{L-M}{ \xi_l} ) }
                        \right]
                   }
                {
                     \sum_{l=0}^{\infty} (2l+1)
\exp{ \left[  - \frac{l(l+1)}{2y} \right] }
                 }, \label{scalepericorr}
\end{equation}
where $ \xi_l $ is the correlation length of the mode $l,$
{\it i.e.}
\begin{equation}
\xi_l = \frac{J}{(l+1)T}. \label{pericl}
\end{equation}
For the open boundary condition,
the correlation function behaves as follows:
\begin{equation}
\left\langle
\left( \frac{ S_{1}^{z}}{S} \right)
\left( \frac{ S_{1+M}^{z}}{S} \right)
\right\rangle^{ \rm { open } }
 \sim f^{ \rm { open } } (M) \quad \mbox{ for} {}^{ \forall } S .\label{ocor}
\end{equation}
Using eq.(\ref{opencorr}),
we have
\begin{equation}
f^{  \rm { open } } (M) \equiv \frac{1}{3} \exp{ ( - M / \xi )} ,
\label{scaleopencorr}
\end{equation}
where the correlation length $\xi$ is $J/T.$

\section*{Acknowledgments}

The authors are grateful to Dr.Nishimori for providing TITPACK version 2.
We also would like to thank Drs. Kaburagi and Tonegawa for KOBEPACK version
1.0.

This work was supported in part by
Grant-in-Aid for Scientific Research on
Priority Areas,
"Molecular Magnetism" (Area No 228),
from the Ministry of Education, Science and Culture, Japan.

\pagebreak
%
%
%
%
%
\eqreset
\renewcommand{\thesection}{Appendix.}
\section{The Correlation Function at the Ground State}
\renewcommand{\theequation}{A.\arabic{equation}}
We show here the derivation of the correlation function (\ref{2.3}).

It is apparent that the Heisenberg ferromagnet has $2LS+1$ number of
degenerate ground states:
\begin{equation}\label{gstates}
\left\{
\left| S_{\rm total}=LS,S^z_{\rm total}=LS-l \right\rangle
\biggm|
l=0,1,\ldots,2LS
\right\}.
\end{equation}
Hereafter we express these states just as $\left| LS-l \right\rangle$
for brevity.
One of the ground states is given by
\begin{equation}\label{allup}
\left| LS \right\rangle
\equiv \bigotimes_{i=1}^L \left| S_i^z=S\right\rangle.
\end{equation}
We can thereby construct the rest of the ground states as
\begin{equation}\label{down}
\left| LS-l \right\rangle
\equiv T_l^-\left| LS \right\rangle
\end{equation}
for $l=0,1,\ldots,2LS$, where the lowering operators $T_l^-$ are defined by
\begin{equation}\label{lower}
T_l^-\equiv\sqrt{\frac{(2LS-l)!}{l!(2LS)!}}\left(S^-_{\rm total}\right)^l
\end{equation}
with
\begin{equation}\label{ladder}
S^-_{\rm total}
\equiv \sum_{i=1}^L S_i^-
=\sum_{i=1}^L \left(S_i^x - {\rm i} S_i^y\right).
\end{equation}
The coefficient in the right-hand side of (\ref{lower}) is added for the
normalization of the states (\ref{down}).

It is easy to see that the state (\ref{allup}) is an eigenstate of the
correlation operator ${\qbSpin}_{1}\cdot {\qbSpin}_{1+r}$:
\begin{equation}\label{sproup}
{\qbSpin}_{1}\cdot {\qbSpin}_{1+r}\left| LS \right\rangle
=S^2\left| LS \right\rangle.
\end{equation}
Now we show that all the states (\ref{down}) are eigenstates of the operator
${\qbSpin}_{1}\cdot {\qbSpin}_{1+r}$ as well.
For this purpose, we first observe the following commutation relation:
\begin{eqnarray}\label{comm}
\lefteqn{
\left[S^-_{\rm total},{\qbSpin}_{1}\cdot {\qbSpin}_{1+r}\right]
}
\nonumber\\
&=&\frac{1}{2}\left[S^-_1+S^-_{1+r},
S^+_1S^-_{1+r}+S^-_1S^+_{1+r}+2S^z_1S^z_{i+r}\right]=0.
\end{eqnarray}
Thereby we have
\begin{equation}\label{comm2}
\left[T_l^-,{\qbSpin}_{1}\cdot {\qbSpin}_{1+r}\right]=0
\end{equation}
for $l=0,1,\ldots,2LS$.
After obtaining these commutation relations, we easily see the following:
\begin{eqnarray}\label{corr}
{\qbSpin}_{1}\cdot {\qbSpin}_{1+r}\left|LS-l\right\rangle
&=&{\qbSpin}_{1}\cdot {\qbSpin}_{1+r} T_l^-\left|LS\right\rangle
\nonumber\\
&=&T_l^-{\qbSpin}_{1}\cdot {\qbSpin}_{1+r}\left|LS\right\rangle
\nonumber\\
&=&S^2T_l^-\left|LS\right\rangle
\nonumber\\
&=&S^2\left|LS-l\right\rangle
\end{eqnarray}
for $l=0,1,\ldots,2LS$.

Thus an arbitrary linear combination of the ground states (\ref{gstates}),
\begin{equation}\label{combi}
\left|\psi_{\rm gs}\right\rangle
=\sum_{l=0}^{2LS}c_l\left| LS-l \right\rangle
\qquad\mbox{with}\quad
\sum_{l=0}^{2LS}\left|c_l\right|^2=1,
\end{equation}
gives the expectation value
\begin{equation}\label{expect}
\left\langle\psi_{\rm gs}\left|
{\qbSpin}_{1}\cdot {\qbSpin}_{1+r}
\right|\psi_{\rm gs}\right\rangle=S^2,
\end{equation}
or eq.\ (\ref{2.3}) in the text.

\pagebreak

\pagebreak
{\bf Figure Captions}

\begin{enumerate}
\item
The temperature dependence of $ \chi T^2 /J  $ for
$ J/(TL) = $ 0.5, 1.0, 1.5, 2.0, 2.5, 3.0 and
5.0 in the classical Heisenberg chain
with the periodic boundary condition.
As $T/J \rightarrow 0 ,$
all the data approach to the values (crosses)
which are given by
the finite-size scaling function (\ref{finite_scale}) with (\ref{3.2.1}).
The solid lines are guides for the eye.
\label{f1}

\item
The temperature dependence of $ \chi T^2 /J  $ for
$ J/(TL) = $ 0.5, 1.0, 1.5, 2.0, 2.5, 3.0 and 5.0
in the classical Heisenberg chain
with the open boundary condition.
As $T/J \rightarrow 0 ,$
all the data approach to the values (crosses)
which are given by
the finite-size scaling function (\ref{open}) with (\ref{3.2.1}).
The solid lines are guides for the eye.
\label{f5}

\item
The $ \sqrt{T/J} $ dependence of $ \chi T^2 /J  $ for
$ J/(TL)$  =  0.5, 1.0, 1.5, 2.0, 2.5, 3.0 and 5.0
in the case of
the $S=1/2$ ferromagnetic Heisenberg chain
with the periodic boundary condition.
We used the systems of size up to $L$ = 14.
Each solid line indicates the extrapolating function
based on the three points nearest to the ordinate.
\label{f2}

\item
The $ \sqrt{T/J} $ dependence of $ \chi T^2 /J  $ for
$ J/(TL) = $ 0.5, 1.0, 1.5, 2.0, 2.5, 3.0 and 5.0
in the case of
the $S=1$ ferromagnetic Heisenberg chain
with the periodic boundary condition.
We used the systems of size up to $L$ = 8.
Each solid line indicates the extrapolating function
based on the three points nearest to the ordinate.
\label{f3}

\item
The finite-size scaling function
$ \chi T^2 /J  =\tilde{\chi}^{\rm peri}(J/TL)$
of the ferromagnetic Heisenberg chain with the periodic boundary condition,
(\ref{finite_scale}).
The solid curve indicates the scaling function  $ {\tilde \chi}^{\rm peri} $
analytically obtained for $S= \infty.$
The numerical data,
which are extrapolated in Figs.\ref{f2} and \ref{f3},
are plotted with circles for $S=1/2$ and with crosses for $S=1.$
\label{f4}

\item
The $ \sqrt{T/J} $ dependence of $ \chi T^2 /J  $ for
$ J/(TL)$  =  0.5, 1.0, 1.5, 2.0, 2.5, 3.0 and 5.0
in the case of
the $S=1/2$ ferromagnetic Heisenberg chain
with the open boundary condition.
We used the systems of size up to $L$ = 14.
Each solid line indicates the extrapolating function
based on the three points nearest to the ordinate.
\label{f6}

\item
The $ \sqrt{T/J} $ dependence of $ \chi T^2 /J  $ for
$ J/(TL) = $ 0.5, 1.0, 1.5, 2.0, 2.5, 3.0 and 5.0
in the case of
the $S=1$ ferromagnetic Heisenberg chain
with the open boundary condition.
We used the systems of size up to $L$ = 8.
Each solid line indicates the extrapolating function
based on the three points nearest to the ordinate.
\label{f7}

\item
The finite-size scaling function
$ \chi T^2 /J  =\tilde{\chi}^{\rm open}(J/TL)$
of the ferromagnetic Heisenberg chain with the open boundary condition,
(\ref{open}).
The solid curve indicates the scaling function  $ {\tilde \chi}^{\rm open} $
analytically obtained for $S= \infty.$
The numerical data,
which are extrapolated in Figs.\ref{f6} and \ref{f7},
are plotted with circles for $S=1/2$ and with crosses for $S=1.$
\label{f8}

\end{enumerate}
\end{document}